\documentclass{ws-ijgmmp}

\begin{document}

\markboth{Authors' Names}
{Instructions for Typing Manuscripts (Paper's Title)}

%
\catchline{}{}{}{}{}
%


\newcommand{\mL}{\mathcal L}
\newcommand{\mP}{\mathcal P}
\newcommand{\mR}{\mathcal R}


\newcommand{\al}{\alpha}
\newcommand{\be}{\beta}
\newcommand{\de}{\delta}
\newcommand{\la}{\lambda}
\newcommand{\ep}{\epsilon}
\newcommand{\vp}{\varphi}


\newcommand{\Ga}{\Gamma}


\newcommand{\kC}{\mathfrak{C}}
\newcommand{\kB}{\mathfrak{B}}


\title{Extended Theories of Gravitation}

\author{Lorenzo Fatibene}

\address{Department of Mathematics, University of Turin, Via Carlo Alberto 10\\
Turin, 10123, Italy\\
INFN - Sezione Torino\\
\email{lorenzo.fatibene@unito.it} }

\author{Simon Garruto}

\address{Department of Mathematics, University of Turin, Via Carlo Alberto 10\\
Turin, 10123, Italy\\
INFN - Sezione Torino\\
\email{simon.garruto@unito.it} }

\maketitle

\begin{history}
\received{(Day Month Year)}
\revised{(Day Month Year)}
\end{history}

\begin{abstract}
In this paper we shall review the equivalence between Palatini$-f(\mR)$ theories and Brans-Dicke (BD) theories at
the level of action principles. We shall define the Helmholtz Lagrangian associated to Palatini$-f(\mR)$ theory and we will define some transformations which will be useful to recover Einstein frame and Brans-Dicke frame.
We shall see an explicit example of matter field and we will discuss how the conformal factor affects the physical quantities.
\end{abstract}


\section{Introduction}

In 1972 Ehlers, Pirani and Schild (EPS) proposed an axiomatic framework for relativistic theories in which instead of assuming a geometric structure on a differentiable manifold $M$, they assumed worldlines of particles and lightrays as primitive objects defined on $M$ itself. They showed how it is possible to define a geometric structure on $M$ assuming only these two primitive objects. See~\cite{EPS}.

Once two congruences of worldlines for particles $(\mP)$ and lightrays $(\mL)$ on $M$ have been defined we can define a {\it conformal class of metrics} $\kC = [g]$. Two metrics $g$ and $\tilde g$ on $M$ are equivalent iff there exists a positive scalar field $\vp(x)$ such that $\tilde g = \vp(x) g.$

Then has been proven that free fall of particle is described by a {\it projective class} $\kB = [\tilde\Ga]$ of connections. See~\cite{OCTGM}. Two connections $\tilde\Ga$ and $\Ga$ are said to be equivalent iff:

\begin{equation}\label{eq:congConn}
\tilde\Ga^\al_{\be\mu} = \Ga^\al_{\be\mu} + \de_{(\be}^\al V_{\mu)}
\end{equation}
for some covector $V_\mu$.

It is possible to see that $\tilde\Ga$ and $\Ga$ share the same geodesic trajectories for any convector $V_\mu$. In this case we say that $\tilde\Ga$ and $\Ga$ are {\it projectively equivalent}.

Since both lightrays and mass particles feel the gravitational field we need a compatibility condition between the conformal class $\kC$ associated to light cones and the projective class $\kB$ associated to free fall. In other words we have to assume that lightlike geodesics are a proper subset of all geodesic trajectories of $\tilde\Ga$.

In view of EPS-compatibility condition we can see that fixed a representative $\tilde\Ga$ of the projective structure $\kB$ there always exists a unique covector $A = A_\mu dx^\mu$ such that:

\begin{equation}
\tilde\nabla g = 2A \otimes g
\end{equation}
where $\tilde\nabla$ is the covariant derivative with respect to $\tilde\Ga$. See~\cite{EEHPFGR}

Equivalently, it is possibile to see that the relation between the representative $g$ of the conformal structure and the representative $\tilde\Ga$ of the projective structure has to be:

\begin{equation}
\tilde\Ga^\al_{\be\mu} = \{g\}^\al_{\be\mu} + (g^{\al\ep}g_{\be\mu} - 2 \de_{(\be}^\al \de^\ep_{\mu)})A_\ep
\end{equation}
where $\{g\}^\al_{\be\mu}$ are the Christoffel of $g$.

In view of EPS framework we will assume a representative $g$ of the conformal class $\kC$ (and two representatives differ for their measurement protocols) and a (torsionless) connection $\tilde\Ga$ which is related to free fall of test particles.

Then dynamics will determine a relation between these two structures. If dynamics enforces EPS-compatibility then the theory is called an {\it extended theory of gravitation (ETG)} and EPS framework is implemented in the field theory. An example of dynamics which implements EPS framework is the class of Palatini$-f(\mR)$ theories.

In Palatini$-f(\mR)$ theories we will assume a metric $g$ and a (torsionless) connection $\tilde\Ga$ as independent fields and we can use the following Lagrangian:

\begin{equation}\label{eq:fRLagr}
L = \sqrt g f(\mR)
\end{equation}
where $\mR$ is the trace of the Ricci tensor $\tilde R_{\be\nu}$ of $\tilde\Ga$ with respect to $g$, namely:

\begin{equation}
\mR = g^{\be\nu} \tilde R_{\be\nu}.
\end{equation}

If we have some matter field in the model (as the electromagnetic field) we can add a term in the Lagrangian \eqref{eq:fRLagr}:

\begin{equation}
L = \sqrt g f(\mR) + \mL_{Mat}(g, \psi)
\end{equation}
where we have denoted  the set of matter fields by $\psi^i$. 

This means that the phase space can be represented by local coordinates $(x^\mu, g_{\mu\nu}, \tilde\Ga^\la_{\mu\be}, \psi^i)$.

Palatini$-f(\mR)$ equation of motions are:

\begin{equation}\label{eq:fREq}
\begin{cases}
f^\prime(\mR) \tilde R_{(\be\nu)} - \frac{1}{2} f(\mR) g_{\be\nu} = T_{\be\nu} \\
\tilde\nabla_\al (\sqrt g f^\prime(\mR)) g^{\be\nu}) = 0 \\
E_i = 0
\end{cases}
\end{equation}
where $E_i = 0$ are the matter field. We can define a conformal factor $\vp = f^\prime$ and we can define another metric:

\begin{equation}
\tilde g_{\be\nu} = \vp  g_{\be\nu}
\end{equation}
and the second equation of \eqref{eq:fREq} is equivalent to:

\begin{equation}
\tilde\Ga^\la_{\be\nu} = \{ \tilde g \}^\la_{\be\nu}
\end{equation}
or in other words the connection $\tilde\Ga$ is the Levi Civita connection of $\tilde g$.

If we trace the first equation of motion with respect to $g$ we obtain the so called {\it master equation} (where the spacetime dimension has been fixed equal to four):

\begin{equation}
f^\prime(\mR) \mR - 2 f(\mR) = T
\end{equation}
with $T = g^{\be\nu} T_{\be\nu}$.

The master equation is an algebraic equation (not differential) for $\mR$. 

For any given (regular enough) function $f(\mR)$ we can define the conformal factor as done before. This operation can be seen as a map:

\begin{equation}
\mR \mapsto \vp = f^\prime(\mR)
\end{equation}
and if this map is invertible we can define the function $\mR = r(\vp)$ which will be useful later.

\section{Helmholtz Lagrangian}

In this section we will introduce the so called {\it Helmholtz Lagrangian}. It is defined as follows:

\begin{equation}
L_H = \sqrt g [
\vp \mR - \vp r(\vp) + f(r(\vp)) 
]+ \mL_{Mat}(g,\psi)
\end{equation}
where we have considered $(g_{\mu\nu}, \tilde\Ga^\la_{\mu\nu}, \vp, \psi^i)$ as independent fields. Equations of this Lagrangian are:

\begin{equation}
\begin{cases}
\vp \tilde R_{\be\nu} = T_{\mu\nu} + \frac{1}{2} f(r(\vp))g_{\be\nu} \\
\tilde\nabla_\la (\sqrt g \vp g^{\mu\nu}) = 0 \\
\mR = r(\vp)\\
E_i = 0
\end{cases}
\end{equation}

It is easy to see that these equations are equivalent to Palatini$-f(\mR)$ equation~\eqref{eq:fREq} together with  the definition of the conformal factor $\vp = f^\prime(\mR)$.

This means that the correspondence:

\begin{equation}
(g_{\mu\nu}, \tilde\Ga^\la_{\mu\nu} \psi^i) \longleftrightarrow (g_{\mu\nu}, \tilde\Ga^\la_{\mu\nu}, \vp, \psi^i)
\end{equation}
defined by $\vp = f^\prime(\mR)$ (or, equivalently, its inverse $\mR = r(\vp)$) sends solutions into solutions and it is 1-to-1.

We will call the choice of $(x^\mu, g_{\mu\nu}, \vp, \tilde\Ga^\la_{\mu\nu}, \psi^i)$ as independent fields the {\it Helmholtz frame}.

\section{Change of Frames}

In this section we will use the freedom allowed on the EPS framework in choosing the representatives of the conformal class $\kC$ and of the projective class $\kB$ in order to define other frames and discuss if these frames define equivalent theories.

Let us begin transforming the metric in the Helmholtz Lagrangian. We shall call the frame described with the coordinates $(x^\mu, \tilde g_{\mu\nu}, \tilde\Ga^\la_{\mu\nu}, \vp, \psi^i)$ the {\it Einstein Frame}, where $\tilde g_{\mu\nu} = \vp g_{\mu\nu}$ and the Helmholtz Lagrangian becomes:

\begin{equation}
L_E(\tilde g_{\mu\nu}, \tilde\Ga^\la_{\mu\nu}, \vp, \psi^i) =
\sqrt{\tilde g} (\tilde g^{\mu\nu} \tilde R_{\mu\nu} + \vp^{-2}(f(r(\vp)) - \vp r(\vp))) + \mL_{Mat}(\vp^{-1}\tilde g, \psi)
\end{equation}
which is a standard Palatini GR with an additional matter field $\vp$.

Since the transformation $\tilde g_{\mu\nu} = \vp g_{\mu\nu}$ is invertible then $L_E$ will give equations which will be equivalent to the equation for Helmholtz Lagrangian.

Now we can transform the connection. We shall call the frame described with the coordinates $(x^\mu, g_{\mu\nu}, \Ga^\la_{\mu\nu}, \vp, \psi^i)$ the {\it Brans-Dicke frame}, where we have defined:

\begin{equation}
\Ga^\la_{\mu\nu} = \tilde \Ga^\la_{\mu\nu}+ \frac{1}{2}\left(g^{\al \ep} g_{\mu\nu} - 2 \de_{(\mu}^\al \de_{\nu)}^\ep
\right) \mathop{\nabla_\ep}^{\ast} \vp
\end{equation}
where $\displaystyle \mathop{\nabla_\ep}^{\ast}$ means that the covariant derivative is independent of any connection.

The new Lagrangian is:

\begin{equation}\label{eq:BDLagr}
L_{BD} (g, \Ga, \vp, \psi) = L_H(g, \tilde\Ga(g, \Ga, j^1\vp), \vp, \psi).
\end{equation}

It can be shown that equations of motion of Lagrangian \eqref{eq:BDLagr} are equivalent to equation of motion of Helmholtz Lagrangian equations. See~\cite{EXTGRA}.

\section{Conclusions and perspectives}

We have defined several frames and we have discussed the equivalence among them at the Lagrangian level. Furthermore, the Brans-Dicke frame provided a Palatini version of Brans-Dicke theory. Indeed if we promote the connection as independent field in the Brans-Dicke Lagrangian, equation of motions will not be equivalent to the original version.

If we couple our Lagrangians with a scalar field like the Klein-Gordon field, we can define and action of the conformal field over the Klein-Gordon field itself as follows:

\begin{equation}
\Phi \mapsto \tilde \Phi = \vp^{-1/2} \Phi
\end{equation}
where $\Phi$ is the Klein-Gordon field. This transformation defines another Lagrangian with a ``variable'' mass $\tilde m$ which is a function of the conformal factor $\vp$.

Since the spectral line of the stars depend on the electric charge and on the mass of the particle we would like to study how they are affected by the conformal factor itself. Further investigations are needed in this way.

\section{Acknowledgments}

This paper is dedicated to the memory of Mauro Francaviglia.

We acknowledge the contribution of INFN (Iniziativa Specifica QGSKY), the local research project {\it Metodi Geometrici in Fisica Matematica e Applicazioni} (2015) of Dipartimento di Matematica of University of Torino (Italy). This paper is also supported by INdAM-GNFM.

\end{document}